\documentclass[prx,aps,showpacs,twocolumn,preprintnumbers,
amsmath,amssymb,superscriptaddress,longbibliography,nofootinbib]{revtex4-2}
\usepackage[utf8]{inputenc}
\usepackage[T1]{fontenc}
\usepackage{amsmath}
\usepackage{tikz}
\usepackage{lipsum}
\usepackage{physics}
\usepackage{comment}
\usepackage{amsfonts}
\usepackage{booktabs}
\usepackage{xspace}
\usepackage[toc,page]{appendix}

\usepackage[breaklinks]{hyperref}
\hypersetup{colorlinks,linkcolor={blue},citecolor={blue},urlcolor={red}}
\usepackage{natbib}
\usepackage{float}

\newcommand{\mpl}{Max Planck Institute for the Science of Light, Erlangen, Germany.}
\newcommand{\theseus}{\textsc{Theseus} \xspace}
\newcommand{\pytheus}{\textsc{PyTheus} \xspace}
\newcommand{\melvin}{\textsc{Melvin} \xspace}

\begin{document}

\title{Deep Quantum Graph Dreaming:\\Deciphering Neural Network Insights into Quantum Experiments}

\author{Tareq Jaouni}%
\email{tjaou104@uottawa.ca}
\affiliation{\mpl}
\affiliation{Nexus for Quantum Technologies, University of Ottawa, K1N 6N5, ON, Ottawa, Canada.}%

\author{Sören Arlt}
\affiliation{\mpl}

\author{Carlos Ruiz-Gonzalez}
\affiliation{\mpl}

\author{Ebrahim Karimi}%
\affiliation{\mpl}%
\affiliation{Nexus for Quantum Technologies, University of Ottawa, K1N 6N5, ON, Ottawa, Canada.}%

\author{Xuemei Gu}%
\affiliation{\mpl}%

\author{Mario Krenn}%
\email{mario.krenn@mpl.mpg.de}
\affiliation{\mpl}%

\begin{abstract}

Despite their promise to facilitate new scientific discoveries, the opaqueness of neural networks presents a challenge in interpreting the logic behind their findings. Here, we use a eXplainable-AI (XAI) technique called \textit{inception} or \textit{deep dreaming}, which has been invented in machine learning for computer vision. We use this technique to explore what neural networks learn about quantum optics experiments. Our story begins by training deep neural networks on the properties of quantum systems. Once trained, we "invert" the neural network -- effectively asking how it imagines a quantum system with a specific property, and how it would continuously modify the quantum system to change a property. We find that the network can shift the initial distribution of properties of the quantum system, and we can conceptualize the learned strategies of the neural network. Interestingly, we find that, in the first layers, the neural network identifies simple properties, while in the deeper ones, it can identify complex quantum structures and even quantum entanglement. This is in reminiscence of long-understood properties known in computer vision, which we now identify in a complex natural science task. Our approach could be useful in a more interpretable way to develop new advanced AI-based scientific discovery techniques in quantum physics.

\end{abstract}


\maketitle


\section{Introduction}

Neural networks have been demonstrably promising towards solving various tasks in quantum science~\cite{modernMLQuantum, reviewMLQuantum, PhysicsAndML}. One notorious frustration concerning neural networks, however, lays in their inscrutability: modern architectures often contain millions of trainable parameters, and it is not readily apparent what role that they each play in the network's prediction. We may, therefore, inquire about what learned concepts from the data that the network utilizes to formulate its prediction, an important prerequisite in achieving scientific understanding~\cite{Krenn_2022_Scientific}. This has since motivated the development of eXplainable-AI (XAI), which interprets how the network comes up with its solutions~\cite{doran_2017_what, tjoa_survey_2021, Burkart_2021, samek2017explainable}. These developments have spurred physicists to address the problem of interpretability, resulting in the rediscovery of long-standing physics concepts~\cite{iten_discovering_2020, siameseNetworks}, the identification of phase transitions in quantum many-body physics~\cite{Dawid_2021, Dawid_2020, K_ming_2021, anotherPhaseOne}, the compression of many-body quantum systems~\cite{Rocchetto_2018},  and the study on the relationship between quantum systems and their entanglement properties~\cite{flam-shepherd_learning_2022, frohnert2023explainable}. 

Here, we apply neural networks in the design of quantum optical experiments. The growing complexity of quantum information tasks has since motivated the design of computational methods capable of navigating the vast combintorical space of possible experimental designs that involve unintuitive phenomena~\cite{Krenn_2020}. To this end, scientists have developed automated design and machine learning routines \cite{melvin}, including some that leverage genetic algorithms \cite{genetic1, genetic2}, active learning approaches \cite{valcarce2023automated} and the optimization of parameterized quantum circuits~\cite{parametric,krenn_conceptual_2021,ruiz-gonzalez_digital_2022}. One may inquire if we may be able to learn new physics from the discoveries made by such algorithms. For instance, the computer algorithm \melvin~\cite{melvin}, which topologically searches for arrangements of optical elements, has led to the discovery of new concepts such as the generation of entanglement by path identity~\cite{PhysRevLett.118.080401} and the creation of multipartite quantum gates~\cite{multipartiteQGates}. However, the interpretability of these solutions is obfuscated by the stochasticity of the processes that create them as well as the unintuitiveness of their representations. The recent invention of \theseus~\cite{krenn_conceptual_2021}, and its successor \pytheus~\cite{ruiz-gonzalez_digital_2022} addresses this through the topological optimization of highly interpretable, graph-based representation of quantum optical experiments. This has already enabled new scientific discoveries, such as a new form of multi-photon interference~\cite{HALO}, and novel experimental schemes for high-dimensional quantum measurement~\cite{meanKing}. 

To this point, the extraction and generalization of new concepts has largely been confined to analyzing the optimal solutions discovered by these algorithms. However, we may inquire if we can learn more physics by probing the rationale behind the computer's discoveries. Little attention has hitherto been given towards the application of XAI techniques on neural networks trained on quantum experiments, which may allow us to conceptualize what our algorithm has learned. In so doing, we may guide the creation of AI-based design techniques for quantum experiments that are more reliable and interpretable. 

In this work, we present an interpretability tool based on the inceptionism technique in computer vision, better known as Deep Dreaming~\cite{noauthor_inceptionism_2015}. This technique has been applied to iteratively guide the automated design of quantum circuits~\cite{lifshitz2022quantum} and molecules~\cite{shen_deep_2021} towards optimizing a target property; it has also been applied in~\cite{schindler_probing_2017} to verify the reliability of a network trained to classify the entanglement spectra of many-body quantum systems. More importantly, it also lets us visualize what physical insights has the neural network gained from the training data. This lets us better discern the strategies applied throughout automated design processes, as well as to verify physical concepts rediscovered by the network, such as the thermodynamic arrow of time~\cite{seif_machine_2021}.

Here, we adapt this approach to quantum graphs. We train a deep neural network to predict properties of quantum systems, then inverse the training to optimize for a target property. We observe that the inverse training dramatically shifts the initial distribution of properties. We also show that, by visualizing the evolution of quantum graphs during inverse training, we are able to conceptualize the learned strategies applied by the neural network. We probe the network's rationale further by inverse training on the intermediate layers of the network. We find that the network learns to recognize simple features in the first layers and then builds up more complicated structures in later layers. Altogether, we synthesize a complete picture of what the trained neural network sees. We, therefore, posit that our tool may aid the design of more interpretable and reliable computer-inspired schemes to design quantum optics experiments.

\section{Methodology}

\subsection{Graphs and Quantum Experiments}
\begin{figure}[H]
  \centering
  \includegraphics[width=\linewidth]{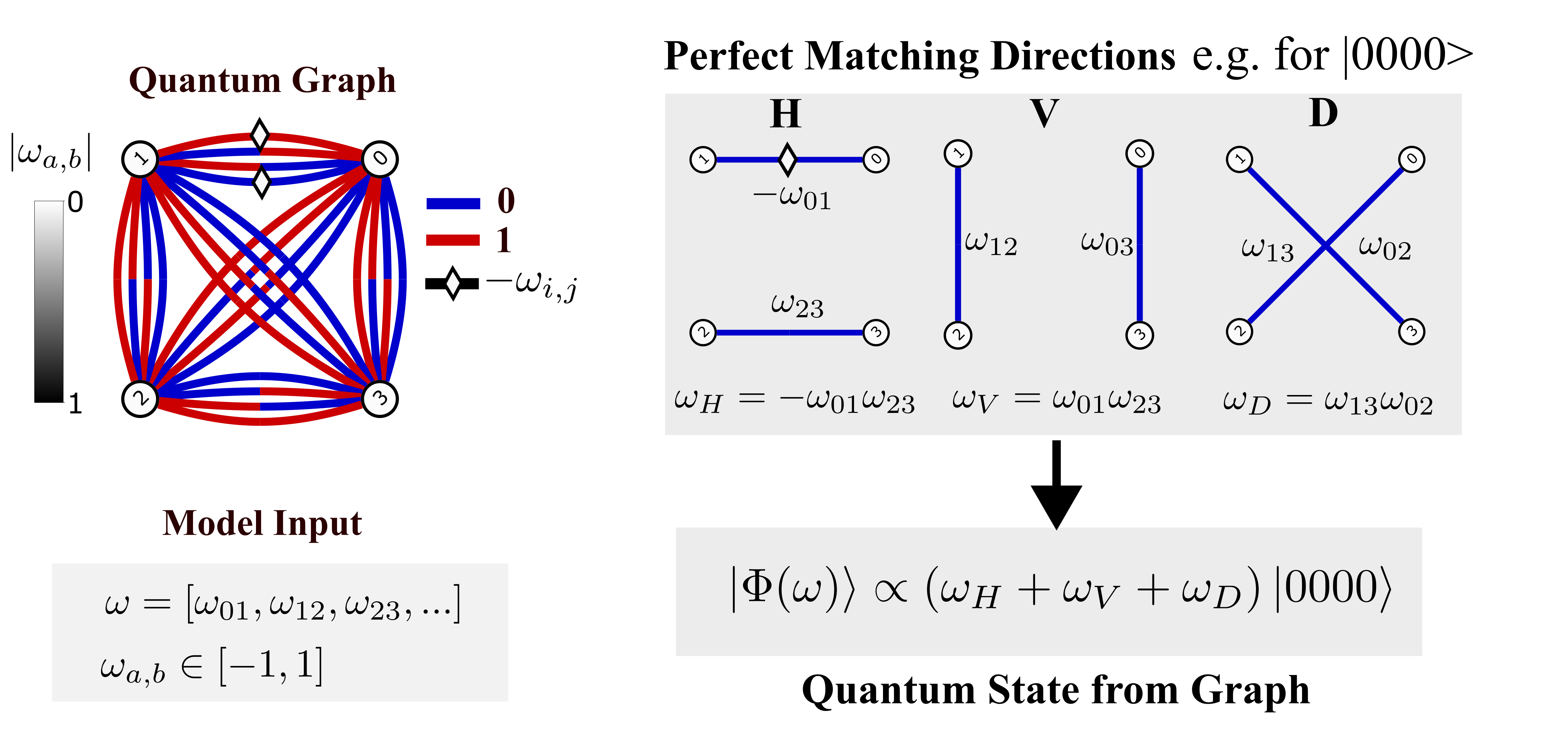}
  \caption{\textbf{Brief overview of quantum graphs}. In this work, we consider \textit{complete} graph representations of two-dimensional, quadripartite quantum graphs. We let $\omega_{a,b}$ denote the weight of the edge connecting vertex $a$ to vertex $b$. The weight's magnitude is indicated by the transparency of the edge and the  presence of a diamond signifies a negative edge weight. The creation of every possible state is conditioned on three possible types of perfect matchings, which are distinguished in terms of their direction. }
  \label{fig:graphbasics}
\end{figure} 
\label{sec:graphs}

\begin{figure*}[t!]
  \centering
  \includegraphics[width=0.9\textwidth]{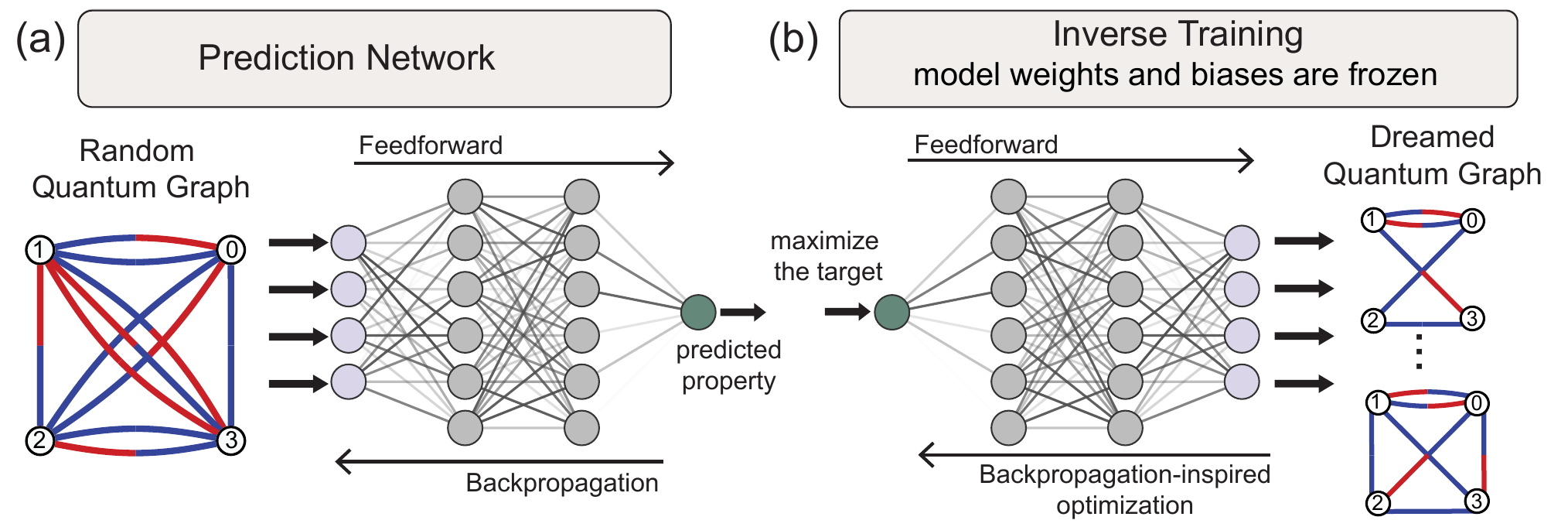}
  \caption{\textbf{Quantum Graph Deep Dreaming.} (a) The weights and biases of a feed-forward neural network are continually updated during training to predict a property such as fidelity of a given input random quantum experiment represented by a graph. (b) In the deep dreaming process, the weights and biases of the network are frozen. The weights of an initial input graph are updated iteratively to maximize the output of the feed-forward network, which gives the network's prediction on the aforementioned property.}
  \label{fig:NNarchitectures}
\end{figure*} 
As developed in~\cite{krenn_quantum_2017,gu_quantum_2019, PhysRevA.99.032338, krenn_conceptual_2021, ruiz-gonzalez_digital_2022}, we may represent quantum optical experiments in terms of colored, weighted, undirected multigraphs. This representation can be extended to integrated photonics~\cite{integratedPhoton1, integratedPhoton2, integratedPhoton3, bao_very-large-scale_2023} and entanglement by path identity~\cite{PhysRevLett.118.080401, Qian_2023, Feng:23}. The vertices of the graph represent photon paths to detectors, whereas edges between any two vertices, $a$ and $b$, indicate correlation between two photon paths. We may assign an amplitude to them by introducing edge weights $\omega_{a,b}$, and we may assign the photons' internal mode number through different edge colorings. We also permit multiple edges between the vertices to indicate the superposition of states. 

Here, we consider graph representations of four-qubit, two-dimensional experiments dealing with state creation. Specifically, we consider graphs with vertices $V=\{0,1,2,3\}$ and mode numbers $0$ and $1$. Each graph, therefore, consists of 24 possible edges with real-valued edge weights between 1 and -1. We may determine the particular quantum state $\ket{\Phi(\omega)}$, where $\Phi(\omega)$ is the graph's weight function defined according to Eq.~(2) in~\cite{ruiz-gonzalez_digital_2022}. We condition the creation of each term in the state on subsets of edges which contains every vertex in the graph exactly once, otherwise known as the perfect matchings (PMs) of the graph. For each term, we can define three possible PMs, each distinguished by their 'directionality', which we show in Figure~\ref{fig:graphbasics}. We obtain the amplitude of the term through the sum of weights of the three perfect matchings, which are themselves determined by the product of edge weights. Applying this procedure for every possible ket in the joint Hilbert space $\mathcal{H} = \mathbb{H}_{2} \otimes \mathbb{H}_{2} \otimes \mathbb{H}_{2} \otimes \mathbb{H}_{2}$, we may obtain the state $\ket{\Phi(\omega)}$.



\subsection{Training}

Figure~\ref{fig:NNarchitectures} illustrates the basic workflow behind the dreaming process. A feed-forward neural network is first trained on the edge weights $\omega$ of a complete, quadripartite, two-dimensional quantum graph in order to make predictions on certain properties of the corresponding quantum state $\ket{\Phi(\omega)}$. We randomly initialize $\omega$ over a uniform distribution $[-1,1]$. The neural network's own weights and biases are optimized for this task via mini-batch gradient descent and the mean squared error (MSE) loss function.


We consider the state fidelity $|\bra{\Phi(\omega)}\ket{\psi}|^2$ with respect to two well-known classes of multipartite entangled states within the joint Hilbert space $\mathcal{H}$. First, the Greenberger-Horne-Zeillinger (GHZ) State~\cite{greenberger_going_2007}, $\ket{\psi} = \ket{\text{GHZ}}$, where
    \begin{equation}
    \ket{\text{GHZ}} = \frac{1}{\sqrt{2}}(\ket{0000} + \ket{1111}),
\end{equation}
and, second, the W-state \cite{cabello_bells_2002}, $\ket{\psi} = \ket{W}$, where 
\begin{equation}
    \ket{\text{W}} = \frac{1}{\sqrt{2}}(\ket{1000} + \ket{0100} +  \ket{0010} + \ket{0001}).
\end{equation}
In addition, we also consider a measure of quantum state entanglement resulting from a graph -- the concurrence~\cite{wooters_concurrence}. Let $A_{1}, A_{2}, A_{3}, A_{4}$ each denote the subsystems of the joint quadripartite Hilbert space to which $\ket{\Phi(\omega)}$ is defined. Then assuming the pure state $\rho = \ket{\Phi(\omega)}\bra{\Phi(\omega)}$, we may write 
\begin{align}
    C(\rho) = \sum_{\mathcal{M}}C_{\mathcal{M}}(\rho) = \sum_{\mathcal{M}}\sqrt{2(\ 1 - \text{tr}(\ \rho^{2}_{M} )\ )\ } 
\end{align}
where $\mathcal{M}$ refers to a bipartition of the subsystem and $\text{tr}(\ \rho^{2}_{\mathcal{M}} )\ $ is the reduced density matrix obtained by tracing out  $\mathcal{M}$. In this work, we train our networks to make predictions on  $\text{tr}(\ \rho^{2}_{\mathcal{M}} )\ $.  Furthermore, for all cases considered, the network is trained on examples with a property value below a threshold of 0.5 to ensure that the network is not memorizing the best solutions in each case. 

\begin{figure*}[!ht]
	\centering
       {\includegraphics[width=\textwidth]{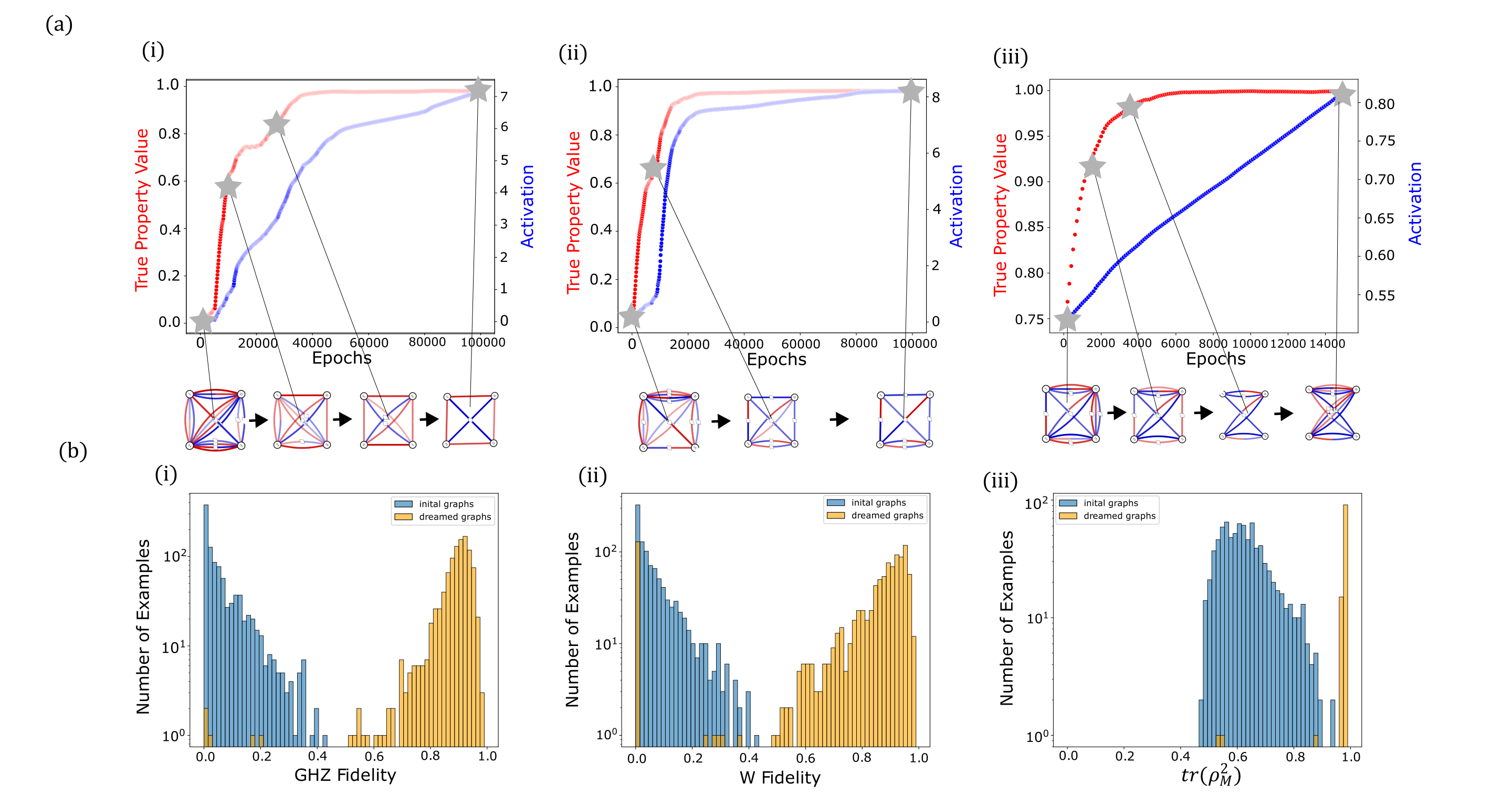}}\!\!
  \caption{\textbf{Dreaming results for on the output layer of different neural network architectures.} (a) Evolution of an input graph's fidelity with respect to (i) the GHZ state and (ii) the W state when dreaming on the $[400^{3}, 10]$ neural network; we also observe (iii) the evolution of an input graph's concurrence when dreaming on the $[800^{7}]$ network. For each case, we show the intermediate steps of the input graphs' evolution to its dreamed counterpart and only show edges whose weights are above a threshold of 0.4. These intermediate steps reveal that, in inverse-training, edges of perfect matchings which do not positively contribute to the target property are mitigated.  (b) Distribution of initial vs. dreamed fidelities with respect to (i) the GHZ state and (ii) the W-state, as well as (iii) the mean value of  tr$(\rho_{\mathcal{M}}^{2})$. We observe that most dreamed examples exceed the upper bound of the original dataset, attesting to our tool's ability to find quantum graphs that are novel to the original dataset. }
  \label{fig: fig2}
\end{figure*} 

Once convergence in the training has been achieved, we then execute the deep dreaming protocol to extract insights on what the neural network has learned. Given an arbitrary input graph, we select a neuron in the trained neural network. Then, we maximize the neuron's activation by updating the input graph via gradient ascent. In this stage, the weights and biases of the neural network are frozen, and we instead optimize for the edge weights of the input graph. At the end of the process, the graph mutates into a configuration which most excites the neuron. However, this may not entirely represent all that the neuron over-interprets from the input graph, as it has been shown in~\cite{nguyen_multifaceted_2016} that individual neurons can be trained to recognize various possible features of the input. Therefore, to uncover all that the neuron sees, we repeat this procedure over multiple different initializations.

\section{Results}

\subsection{Dreaming on the Output Layer}

Towards attaining a general idea of what the neural network has learned about select properties for the quantum state $\ket{\Phi(\omega)}$, we first apply the deep dreaming approach on the output layer. Figure~\ref{fig: fig2}(a) illustrates the mutation of an input graph by applying the deep dreaming approach on a [$400^{3}$,10] (three hidden layers of 400 neurons, one hidden layer of 10 neurons) neural network, which has been trained to predict either the GHZ-state or the W-state fidelity. We also apply this approach on a $[800^{7}]$ neural network architecture, which has been trained to predict the mean value of $\Tr( \rho_{\mathcal{M}}^{2} )$. While dreaming, we task our network to find configurations which maximizes the property value. It should be stressed that, in particular, the optimal configuration that maximizes tr$\overline{( \rho_{\mathcal{M}}^{2} )}$  \textit{minimizes} the concurrence; we, therefore, anticipate the dreamed graph to correspond to a maximally separable state.  

We obtain $\ket{\Phi(\omega)}$ from the reconstructed, mutated graph and recompute its true property value in each step. In all cases, we find that the graph evolves steadily towards the maximum property value. We repeat this procedure for 1000 different quantum graphs and plot the distribution of each graphs' initial versus dreamed fidelities in Figure~\ref{fig: fig2}(b). In all three cases, we observe that the network consistently finds distinct examples with a property value outside the initial distribution's upper bounds. This demonstrates our approach's potential to discover novel quantum graphs which optimizes a specific quantum state property.

\begin{figure*}[!ht]
	\centering
       {\includegraphics[width=\textwidth]{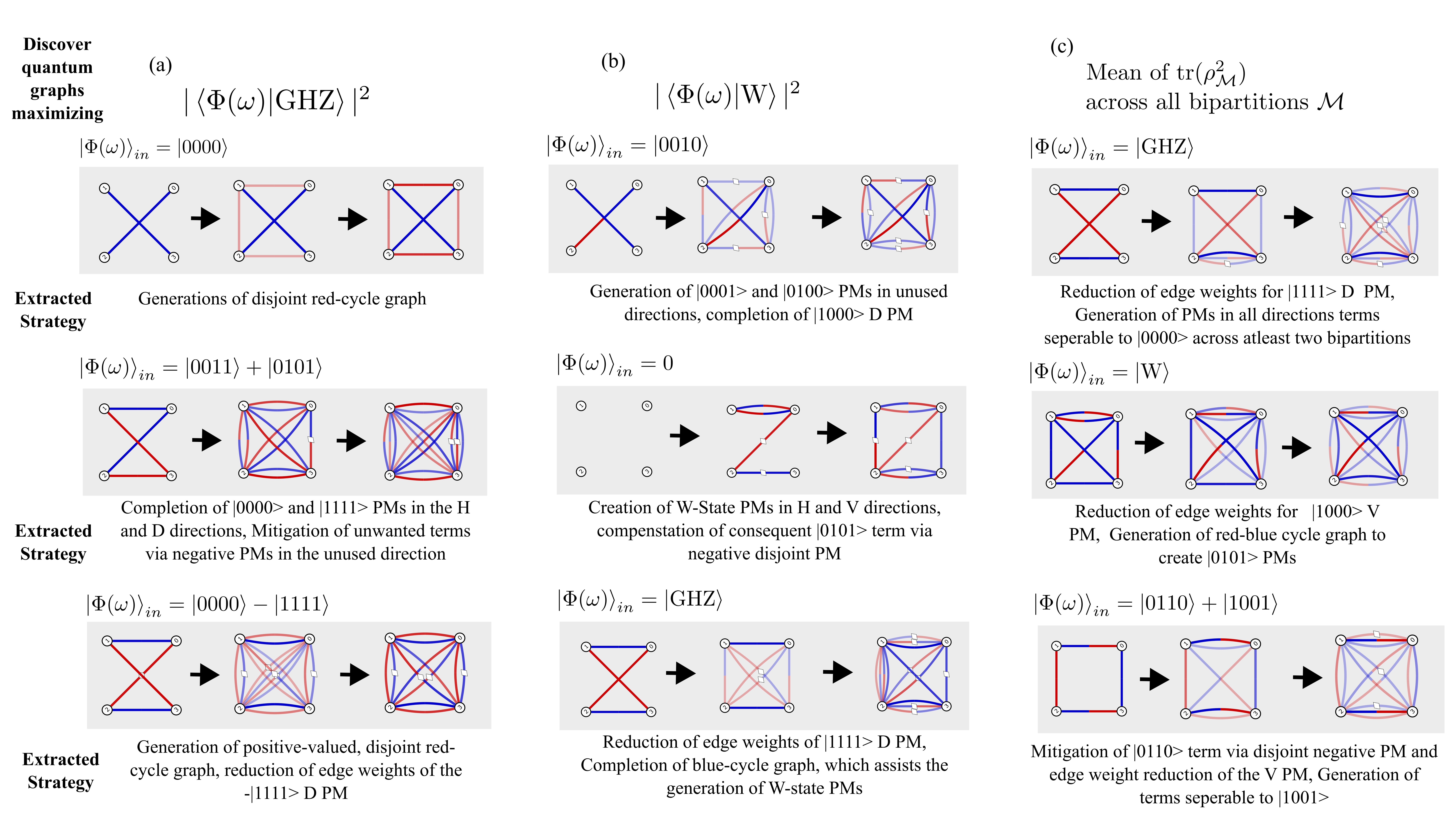}}\!\!
  \caption{\textbf{Extracted strategies from the evolution of certain states when dreaming on the output layer of the neural network.} We discern the strategies employed by the inverse training routine when applied to a network tasked to optimize (a) the GHZ-State Fidelity, (b) the W-State Fidelity, and (c) the mean value of tr$(\rho_{M}^{2})$ by considering several initialisations for each case. For each graph, we only show edges with weights greater than 0.3. We find that the network attempts to construct perfect matchings (PMs) of terms which positively contribute to the property value and whose weights add up to 1. Conversely, we find that the network eliminates unwanted terms by either directly reducing the edge weights of the PM corresponding to that term, or by introducing negative, disjoint perfect matchings of that term.  For (c), we observe that the network 'selects' a term in the initial state to be minimized, then creates terms that are separable across two or more bipartitions with respect to the remaining states.}
  \label{fig: fig2_2}
\end{figure*} 

The intermediate steps of the dreaming process allow us to discern what strategies the neural networks are applying to a given optimization task. In Figure~\ref{fig: fig2_2}, we summarize the evolution of different initial graphs during inverse training for different targets. In Figure~\ref{fig: fig2_2}(a), we observe that the neural network tries to activate the $\ket{0000}$ and $\ket{1111}$ states either by creating perfect matchings (PM) of these terms in unused directions -- the input graph had no PM in that direction previously -- or by completing them with the assistance of an existing PM in some direction, as is seen in particular with the $\ket{\Phi(\omega)} = \ket{0011} + \ket{0101}$ initialization. We note that the dreaming process creates these PMs such that their weights add up to 1.  In circumstances where the initial graph starts with unwanted terms, or when the network unavoidably creates these terms while dreaming, the network attempts to eliminate them either by directly lowering the edge weights' magnitudes or by introducing negative weight PMs in different directions. We see this trend continue when the network is tasked with maximizing the W-state fidelity, as shown in Figure \ref{fig: fig2_2} (b), albeit instead in favouring the activation of the $\ket{1000}, \ket{0100}, \ket{0010},$ and $ \ket{0001}$ states. In Figure~\ref{fig: fig2_2} (c), the network attempts to maximize the separability of the initial, maximally entangled state by first eliminating one term from the initial state via edge-weight minimization or through negative PMs, then creating PMs of additional terms which are separable with respect to the intermediate graph state across two or more bipartitions. 

Through our deep dreaming approach, we have shown that the network learns about creating states through the graph representation in order to consistently achieve optimal values for select properties of the quantum state. We remark that, for each state property,  the network was able to ascertain the configurations which maximizes them while only seeing configurations having property values below 0.50. This strongly suggests that the network is achieving its tasks from physical insights, rather than by memorizing the best examples. 

\subsection{Interpretability of Neural Network Structure}
\begin{figure*}[!ht]
\centering
       {\includegraphics[width=\textwidth]{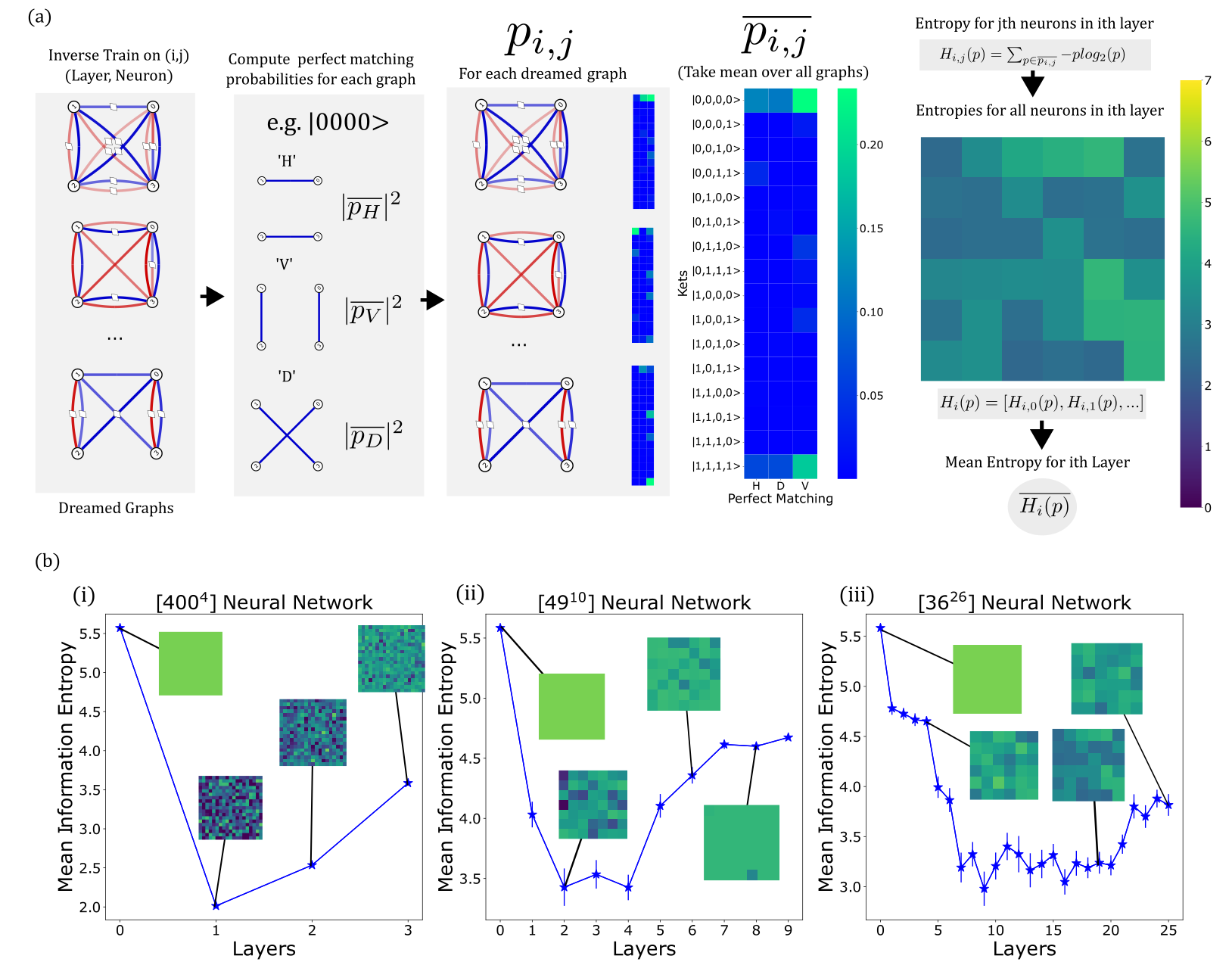}}\!\!
\caption{\textbf{Information entropy throughout each different neural network architecture.} (a) Workflow behind computing the mean information entropy for each layer of the trained neural network. We dream with multiple input graphs on each neuron in the neural network. To account for the diversity of structures that a neuron is interested in seeing, We compute the mean probability amplitudes for every possible perfect matching corresponding to each ket. We thereby observe the overall graph, which the neuron sees best. We may then compute the information entropy of each neuron, $H_{i,j}(p)$, and the mean information entropy of the layer,  $\overline{H_{i,j}(p)}$. This gives us a measure of the complexity of structures seen by the neural network. As conveyed in the different $p_{i,j}$ for each dreamed graph, we note the variety of structures which the network over-interprets; this illustrates the multifactedness of the neurons. (b) Mean information entropy plots for the (i) $[400^{4}]$ (ii) $[49^{10}]$ and (iii) $[36^{26}]$ neural network architectures. A general trend that we may discern in all three cases is that the mean information entropy converges to a minimum in the lower layers and then gradually increases as we go deeper. We may attribute this to the intuition that the network initially learns to recognize simpler structures, then learns increasingly complicated ones as we go deeper within the network.}
 \label{fig: fig4}
\end{figure*}

We apply the deep dreaming approach on the neurons of its hidden layers to gain insight into the neural network's internal model, which generalizes well beyond the training data. We summarize the insights that we extract through our routine in Figure~\ref{fig: fig4}. To showcase the universality of our approach, we consider several different neural network architectures -- the $[400^{4}]$, $[49^{10}]$ and $[36^{26}]$ networks -- that have each been trained to predict the GHZ-state fidelity. For each network, we dream on the $i^{th}$ neuron in the $j^{th}$ hidden layer with 20 input graphs to best capture all of the possible structures exciting the neuron. 

We take particular interest in how the complexity of the dreamed graphs evolves with the network depth. We obtain the greatest amount of information about our quantum graphs by considering all of the different ways, as seen through the graphs' PMs, that a ket is realized. We, therefore, attribute to each dreamed graph a $3 \times 16$ array, $p_{i,j}$, consisting of the probabilities of all possible PMs; through this, we gain insight into the state created by the graph, as well as all PM directions being used to that purpose. As we go deeper into the neural network, we observe that the dreamed graphs activate a greater number of PM directions and kets, which reflects the increasing complexity of structures the neural network has learned to recognize. We also verify the multifaceted nature of the neurons: different input graphs are observed to result in dreamed graphs that recreate different input states. As we see in the third inset of Figure \ref{fig: fig4} (a), the neuron may over-interpret parts of the graph that best creates the $\ket{0000}$ term, or it may  either over-interpret different possible PM directions for $\ket{0000}$, or parts of the graph which instead realize the $\ket{1111}$ term.

We may quantify the complexity of structures recognized throughout the network with the information entropy $H_{i,j}$. We take the mean value of $p_{i,j}$ across all of the dreamed graphs, then use it to compute $H_{i,j}$ through the procedure outlined in Appendix \ref{entropyCalc}. Repeating this procedure across all hidden layer neurons, we may then determine the average entropy observed across the $j^{th}$ layer, which gives us a general metric of the complexity of structures being recognized. We plot the trend of $\overline{H_{i,j}}$ observed across all three neural network architectures in Figure \ref{fig: fig4}(b). Intuitively, we expect that a deep neural network first learns to recognize simple structures, then more abstract features with network depth. Indeed, we observe consistently that, from an initial peak, the information entropy drops to its lowest values at the earlier layers, before gradually increasing near the end of the neural network. This certifies the universal assertion that the network identifies simple features of the input graph, such as edges that form one or two PMs to states, before forming more complicated graphical structures in the deeper layers that features a greater set of PMs. 

\section{Outlook}
In this article, we showcase preliminary results for adapting the deep dreaming approach to quantum optical graphs for deep neural networks on different target quantities. We apply our routine to ascertain the strategies employed by the neural network on its predictive task by dreaming on the output layer and throughout the network. Crucially, we demonstrate that the trained neural network builds a non-trivial model of the quantum state properties produced by a quantum experiment, and we find that the deep dreaming approach does remarkably well in finding novel examples outside of the initial dataset. Lastly, in applying our approach to the hidden layers of the neural network, we find that the network gradually learns to recognize increasingly complicated structures, and that the individual neurons are multifaceted in the possible structures that excites them. In future work, further transparency of the learned representations can be possibly attained by applying regularization techniques such as $\alpha$-norm~\cite{simonyan_deep_2014}, jitter~\cite{noauthor_inceptionism_2015}, or by dreaming on the mean of a set of input graphs~\cite{nguyen_multifaceted_2016} to converge towards more interpretable solutions. Furthermore, we may also find simpler networks on which to dream by applying pruning strategies based on the Lottery Ticket Hypothesis~\cite{frankle2018lottery}. Above all, we may also apply these tools to larger graphs with more dimensions and explore different applications beyond state creation, such as Quantum Measurements and Quantum Communication. 

Thanks to their relative simplicity, the quadripartite graphs have been a good testing case for our inception approach, and the knowledge we extract from them can be used in other systems. Larger graphs and new targets will provide a novel and deeper understanding of quantum optics experiments as well as inspire new research. We foresee that our approach can be used to extend frameworks for automated setup design~\cite{ruiz-gonzalez_digital_2022, melvin, Krenn_2022_Scientific} as well as in generative molecular algorithms~\cite{shen_deep_2021, quantumChem_diffusion} which adapt a surrogate neural network model. Through our approach, we can better decipher what these frameworks have learned about the underlying science, and understand the intermediate strategies taken towards a target configuration. 
\section*{Code Avaliability}
We provide the data featured in this work, as well as the code that executes the deep dreaming protocol, can be found \href{https://github.com/artificial-scientist-lab/deepGraphDreaming }{in this GitHub repository.}

\section*{Acknowledgements}
T.J. and E.K. acknowledge the support of the Canada 
Research Chairs (CRC) and Max Planck-University of Ottawa Centre for Extreme and Quantum Photonics. 

\bibliography{arXiv} 
\clearpage

\section{Appendix}
\subsection{Training Details}
We generate 20 million input-output pairs using the digital discovery framework  \pytheus ~\cite{ruiz-gonzalez_digital_2022}. Each input is a one-dimensional array consisting of the 24 real-valued edge weights which corresponds to the quantum graph, and the output is the property value of the graph's corresponding state, $\ket{\Phi(\omega)}$. The network is then forward-trained on these examples with a mini-batch size of 5000 and a 95:5 train-test split. We utilize a learning rate scheduler that is initially set $1\times10^{-3}$ ($1\times10^{-5}$ for the $[36^{26}]$ training architecture) and is gradually decreased in factors of 0.95 if, after every 25 training epochs, the test MSE does not change significantly. The network is run until convergence in the test MSE is maintained for over 400 training epochs. Table \ref{table1} lists the network architectures considered and the corresponding training results for each network. Between the networks' hidden layers,  We employ the ReLU nonlinear activation function for all of our architectures except for the $[36^{26}]$ network, in which the ELU activation function with $\alpha=0.1$ was used. The networks were encoded using PyTorch \cite{pytorch}, and we employ the Adam optimizer \cite{adam} for both the forward and inverse training steps.


\begin{figure*}[t!]
\centering
       {\includegraphics[width=\textwidth]{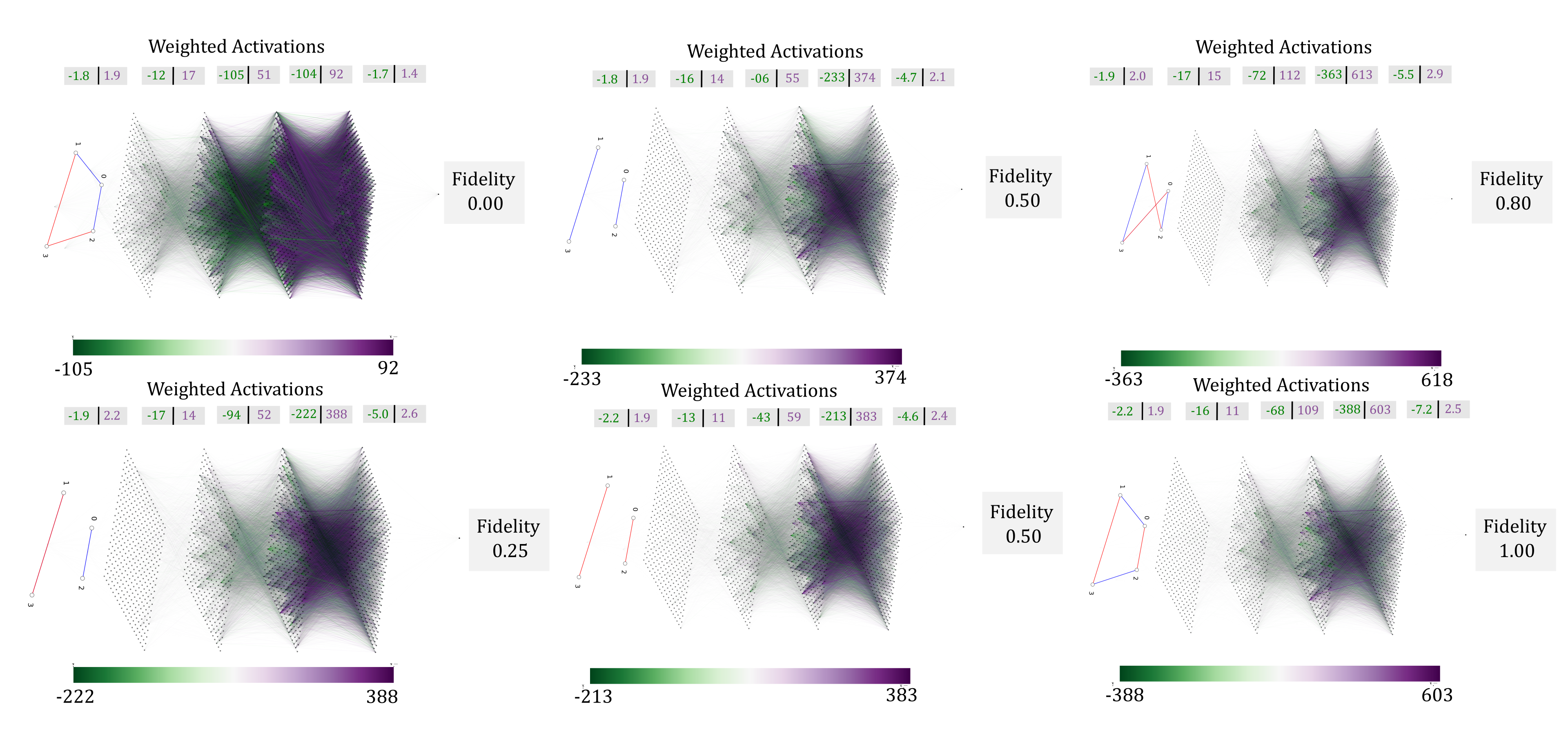}}\!\!
\caption{\textbf{Plots of the \textit{globally normalized} weighted activation for the  $[400^{4}]$ neural network architectures when trained to predict the fidelity of the GHZ state.} The plots are computed by taking the product of the trained neural network weights with the activations of the network when asked to predict $|\braket{\Phi(\omega)}{\text{GHZ}}|^{2}$ for various states $\ket{\Phi(\omega)}$. Towards best visualizing each network, we filter out edges whose weighted activations are below a threshold of 0.05. We find that different perfect matching directions and colourings change which neurons are activated in the intermediate layers, whereas the corresponding state's fidelity changes the magnitude of activations observed throughout the network. The different kinds of neuron activation patterns give us insight into how the neural network conceives its prediction. 
}
  \label{fig: fig_weightedActivationPatterns}
\end{figure*}

We perform inverse training on the $j^{th}$ neuron in layer $i$ each of our trained networks' architecture by employing gradient ascent on random input graphs towards maximizing the neuron's activations. This is done by transferring all of the trained networks' parameters up to and including the $(i-1)^{th}$ layer. The $i^{th}$ layer, which now forms the output layer of this intermediate network, consists solely of neuron $j$. We then sample randomly from a dataset consisting of 1 million input graphs and perform the dreaming routine for 100,000 epochs and a fixed learning rate of $1\times 10^{-4}$. 

\begin{table}
\label{table1}
\caption{\textbf{Training Details of all Neural Network Architectures featured in this work.} Each architecture is listed in the format $[N_{1}, N_{2}, ...N_{i}]$, where $N_{i}$ refers to the number of neurons of the $i^{th}$ hidden layer. }
\begin{tabular}{ccc} 
    NN Architecture &  Test MSE & Training Epochs  \\ \midrule
    $[400^{4}]$  & $2.48\times10^{-6}$ & 2550 \\
    $[800^{7}]$ & $2.95\times 10^{-6}$ &  2300 \\ 
    $[49^{10}]$  & $3.60\times 10^{-4}$  &  2300 \\
    $[36^{26}]$  & $6.24\times 10^{-4}$  & 2700 \\
    $[400^{3}, 10]$   &  
    \shortstack{$4.12\times10^{-6}$ (GHZ)  \\ $6.89\times10^{-6}$ (W)} & 4200 (Both) \\
    \midrule
    \bottomrule

\end{tabular}
\end{table}

\subsection{Neural Network Activation Plots}
Figure~\ref{fig: fig_weightedActivationPatterns} displays the weighted activation patterns for the trained $[400^{4}]$ neural network architecture when tasked to make predictions on quantum graphs over a range of fidelity values. We observe that the activation patterns change depending on the kinds of perfect matchings featured by the input graph. In tandem with our deep dreaming approach, we envision that we can reverse engineer what the neural network is observing in its computation of the GHZ-state fidelity by examining the activation patterns and through knowledge of what each individual neuron is precisely seeing. 

\subsection{Quantifying the Quantum Graph Complexity}
\label{entropyCalc}


The complete Hilbert space $\mathcal{H} = \mathbb{H}_{2} \otimes \mathbb{H}_{2} \otimes \mathbb{H}_{2} \otimes \mathbb{H}_{2}$ on which $\ket{\Phi(\omega)}$ is defined consists of 16 possible states. In the formulation of quantum graphs estabilshed in Section \ref{sec:graphs}, the probability amplitude of any quantum state $\ket{\psi} \in \mathcal{H}$ can be obtained via the weights of the three possible perfect matchings which realize the state. The weights for each perfect matching are obtained as follows,  
\begin{equation}
    p_{\ket{\psi}} = p^{\ket{\psi}}_{H} + p^{\ket{\psi}}_{V} +  p^{\ket{\psi}}_{D}
\end{equation}
where
\begin{align}
p^{\ket{\psi}}_{H} &= |\omega_{01}\omega_{23}|^{2} \\
p^{\ket{\psi}}_{V} &= |\omega_{03}\omega_{12}|^{2} \\
p^{\ket{\psi}}_{D}  &= |\omega_{02}\omega_{13}|^{2} 
\end{align}
and $0 \leq |\omega_{a,b}|^{2} \leq 1$, $a,b  \in \{0,1,2,3\}$ denote the weight of an edge with vertices a,b for the endpoint.

Repeating this procedure for every ket in $\mathcal{H}$, we obtain a $3 \times 16$ array of probabilities $p_{i,j}$ for the $i^{th}$ neuron of the $j^{th}$ hidden layer. We may think of the graphs' complexity in terms of the sum of probabilities that different possible events -- here, the perfect matchings corresponding to each quantum state -- may occur. The overall complexity of structures observed by our neuron can, therefore, be quantified by calculating $\overline{p_{i,j}}$ over all of our representations, and computing the information entropy. 
\begin{equation}
    H_{i,j} =  \sum_{p \in \overline{p_{i,j}}} -p\log_{2}(p)
\end{equation}
We may iterate with this procedure across all of the neurons in the $j^{th}$ layer, yielding an array of information entropies out of which the mean information entropy of the layer, $\overline{H_{j}}$, may be obtained. This gives us an overall measure of the complexity of structures being observed at every point in the neural network. 

\end{document}